\documentclass[twocolumn]{aastex6}
\newcommand{\water}{H$_2$O}

\newcommand{\methane}{CH$_4$}
\newcommand{\kms}{km~s$^{-1}$}
\newcommand{\kp}{$K_\mathrm{P}$}

\newcommand{\pname}{HD~209458~b}

\shorttitle{Combined low- and high-resolution spectroscopy of transiting exoplanets}
\shortauthors{Brogi et al.}
\begin{document}

\title{A framework to combine low- and high-resolution spectroscopy \\  for the atmospheres of transiting exoplanets}

%% Use \author, \affil, plus the \and command to format author and affiliation 
%% information.  If done correctly the peer review system will be able to
%% automatically put the author and affiliation information from the manuscript
%% and save the corresponding author the trouble of entering it by hand.
%%
%% The \affil should be used to document primary affiliations and the
%% \altaffil should be used for secondary affiliations, titles, or email.

%% Authors with the same affiliation can be grouped in a single
\author{M. Brogi\altaffilmark{1,6}, M. Line\altaffilmark{2}, J. Bean\altaffilmark{3}, J.-M. D\'esert\altaffilmark{4}, and H. Schwarz\altaffilmark{5}}
%\affil{American Astronomical Society \\
%2000 Florida Ave., NW, Suite 300 \\
%Washington, DC 20009-1231, USA}

%\author{Butler Burton\altaffilmark{3}}
%\affil{National Radio Astronomy Observatory}

%\affil{TeXnology Inc}

%\author{Julie Steffen\altaffilmark{4}}
%\affil{American Astronomical Society \\
%2000 Florida Ave., NW, Suite 300 \\
%Washington, DC 20009-1231, USA}

%% Use the \and command so offset the last author.
%\and

%\author{Jeff Lewandowski\altaffilmark{5}}
%\affil{IOP Publishing, Washington, DC 20005}


\altaffiltext{1}{Center for Astrophysics and Space Astronomy, University of Colorado at Boulder, Boulder, CO 80309, USA}
\altaffiltext{2}{School of Earth and Space Exploration, Arizona State University, Tempe, AZ 85287, USA}
\altaffiltext{3}{Department of Astronomy \& Astrophysics, University of Chicago, 5640 S Ellis Ave, Chicago, IL 60637, USA}
\altaffiltext{4}{Anton Pannekoek Institute for Astronomy, University of Amsterdam, 1090 GE Amsterdam, The Netherlands}
\altaffiltext{5}{Leiden Observatory, Leiden University, Postbus 9513, 2300 RA Leiden, The Netherlands}
\altaffiltext{6}{NASA Hubble Fellow}

\begin{abstract}
Current observations of the atmospheres of close-in exoplanets are predominantly obtained with two techniques: low-resolution spectroscopy with space telescopes and high-resolution spectroscopy from the ground. Although the observables delivered by the two methods are in principle highly complementary, no attempt has ever been made to combine them, perhaps due to the different modeling approaches that are typically used in their interpretation. Here we present the first combined analysis of previously-published dayside spectra of the exoplanet \pname\ obtained at low resolution with HST/WFC3 and Spitzer/IRAC, and at high resolution with VLT/CRIRES. By utilizing a novel retrieval algorithm capable of computing the joint probability distribution of low- and high-resolution spectra, we obtain tight constraints on the chemical composition of the planet's atmosphere. 
In contrast to the WFC3 data, we do not confidently detect \water\ at high spectral resolution. The retrieved water abundance from the combined analysis deviates by 1.9$\sigma$ from the expectations for a solar-composition atmosphere in chemical equilibrium. Measured relative molecular abundances of CO and H2O strongly favor an oxygen-rich atmosphere (C/O$<1$ at $3.5\sigma$) for the planet when compared to equilibrium calculations including O rainout. From the abundances of the seven molecular species included in this study we constrain the planet metallicity to 0.1-1.0$\times$ the stellar value (1$\sigma$).
This study opens the way to coordinated exoplanet surveys between the flagship ground- and space-based facilities, which ultimately will be crucial for characterizing potentially-habitable planets.
\end{abstract}


\section{Introduction} \label{sec:intro}

Fifteen years of observations with photometry and Low-Dispersion Spectroscopy (LDS) have significantly advanced our understanding of exoplanet atmospheres \citep[e.g.,][]{sin16}. In particular, the Hubble Wide Field Camera 3 (WFC3) has produced robust and repeatable observations, resulting in physically sensible atmospheric composition interpretations. Its band around 1.4\,\micron\ has a strong diagnostic power since it targets major sources of opacity for exoplanet atmospheres (primarily \water, but also \methane, HCN, and NH$_3$). The interpretation of LDS data is also supported by mature retrieval algorithms \citep[e.g.,][]{mad11, ben12, lin13, wal15}. When multiple species overlap, however, ambiguities arise in molecular identification and abundance determinations.

High-Dispersion Spectroscopy (HDS) from the ground has recently emerged as an additional tool to characterize exoplanet atmospheres \citep[e.g.,][]{sne10, sne14, bro12}. HDS is able to distinctly determine the presence of specific molecular species, despite overlap. For emission spectra, absorption and emission are intrinsically discriminated, allowing the direct detection of thermal inversion layers. Due to the large opacity difference between the core and the wings of molecular lines, HDS probes a broad range in atmospheric temperatures and pressures. However, retrieving atmospheric properties (especially absolute molecular abundances) from HDS data is challenging. This is partly due to the loss of the planet+star continuum from self-calibration of the data, and additionally to the lack of robust retrieval algorithms. Inference in HDS data analysis is done through forward modeling, since atmospheric signatures are searched by cross-correlating with template spectra, which combines the signal of tens or hundreds of molecular lines. Not only is this process computationally cumbersome, but also the statistical properties of the resulting cross-correlation functions (CCFs) are far from trivial.

In this paper we present a novel analysis technique to combine the strengths of high-resolution and low-resolution spectroscopy. 
We demonstrate the method on dayside spectroscopic observations of the hot Jupiter \pname\ \citep{hen00,cha00}. As one of the best-characterized transiting exoplanets, it is the ideal target for testing our new method. We aim to show that, when combining LDS and HDS, the sensitivity to a wider range in atmospheric pressures and the ability of HDS to disentangle overlapping molecular features can significantly improve the inference of the chemical compositions of exoplanet atmospheres.

\section{Observations and reanalysis of CRIRES spectra}\label{sec:obs}

\begin{figure}[ht!]
\epsscale{1.25}
\plotone{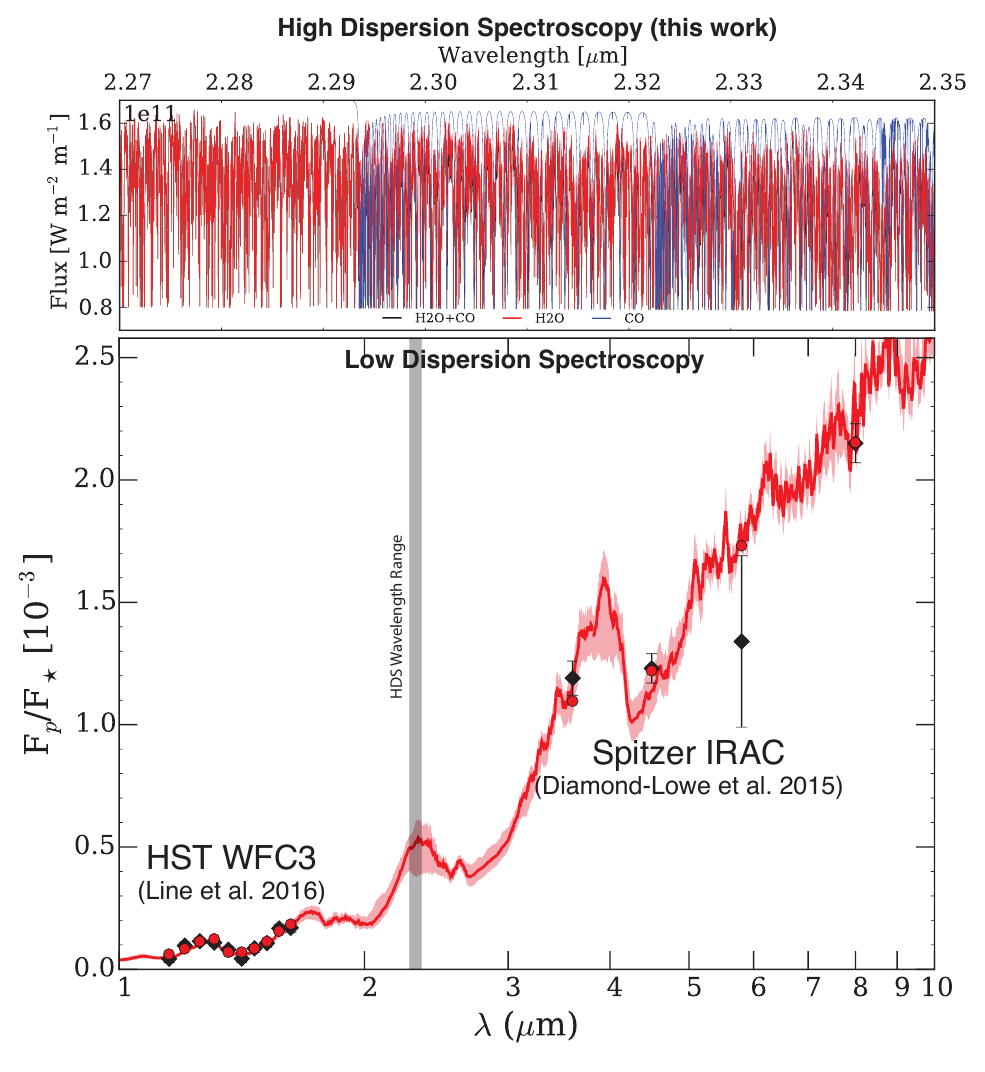}
\caption{Dayside spectrum of \pname. Bottom: LDS data (WFC3+Spitzer, black diamonds), with the best-fitting low-resolution model spectrum and its 1-$\sigma$ uncertainty over-plotted in red. Top: Best-fitting HDS model from this analysis, matching the range of CRIRES 2.3-\micron\ data.}
\label{fig:lds_hds}
\end{figure}

The LDS dataset used for this work is described in \citet{lin16} and shown in Figure~\ref{fig:lds_hds}. It consists of eclipse observations of \pname\ obtained with Hubble/WFC3 in the range 1.125-1.655$\,\mu$m, and with Spitzer/IRAC at $3.6,4.5,5.8,$ and 8.0~\micron\ from \citet{dia14}. 

The HDS data consist of two half nights of dayside spectra taken with VLT/CRIRES in the range 2.287-2.345$\,\mu$m at a resolution of 100,000, and are described in detail in \citet{sch15}. We only include the nodded observations (2 out of 3 sets of data) for which we have a good understanding of the instrument. 
We utilize the same calibrated and extracted spectra as \citet{sch15}, with one important difference. In our current analysis we also estimate the weight of each CRIRES detector based on the injection and retrieval of an artificial signal. The artificial signal is the best-fitting, CO-only model spectrum of \citet{sch15}, scaled to planet/star flux units following their prescriptions, and Doppler-shifted to the planet radial velocity computed with a semi-amplitude of \kp~$=(145.9\pm2.4)$~\kms. The orbital solution is obtained from the literature \citep{knu07, tor08}, assuming a circular orbit.

The scaled and shifted model is injected in the data at 5$\times$ the nominal level, which allows us to achieve a $\mbox{S/N}\ge3$ in all of the four detectors. The injection occurs as early as possible in the analysis, i.e., after the spectra are aligned to the common telluric reference frame. We then remove telluric lines and cross correlate the residual spectra with the model as in \citet{bro14}. The injected artificial signal is retrieved by co-adding the CCFs from each night and each CRIRES detector at the planet radial velocity. We find that all four CRIRES detectors contain significant signal, but their relative weight differs between nights and between detectors, possibly due to varying observing conditions and the unequal density of CO spectral lines with wavelength. When weighting the CCFs between the observed spectra and the model by the square of the measured artificial S/N, i.e., max(CCF)/r.m.s.(CCF), the tentative detection of \citet{sch15} becomes a solid detection of CO in absorption at $\mbox{S/N}=5.3$. We do not proceed to explore their full model grid at this stage, but we utilize the weighting and the analysis described above to test a novel retrieval method. We note that no detection is obtained from models containing \water\ alone. 

\section{Joint LDS+HDS analysis}\label{analysis}
Our aim is to combine the complementary information from LDS and HDS and compute a joint posterior distribution. 
LDS data can be compared to models in a straightforward manner (e.g., via "chi-square"), whereas the signal in HDS data is extracted by cross-correlating with a family of models. We thus need to define an appropriate metric for quantifying the fit quality of a model to HDS data. Since comparing models via cross correlation is computationally expensive, we also need to design an efficient exploration of the parameter space.

\subsection{A metric for the fit quality of high-resolution model spectra}\label{sec:hds_metric}
The cross-correlation function between the data and a model is maximized when both the position and the strength of spectral lines {\it relative} to each other match. However, due to invariance under scaling, maximizing the cross-correlation signal is not equivalent to matching the {\sl absolute} line-to-continuum contrast. This additional constraint can be fulfilled by minimizing the residual cross-correlation signal when each model is subtracted from the data. In other words, the best-fitting model will maximize the cross correlation with the observed spectra and minimize the cross correlation with the model-subtracted data. These two requirements are incorporated into a consistent statistical framework adapted from \citet{bro16}:

\begin{itemize}

\item The model is directly cross-correlated with the data after telluric lines are removed\footnote{ Stellar lines - especially CO lines - are negligible in these CRIRES spectra, and no spurious cross-correlation from the star is detected at \kp~=~0 and at the systemic velocity of HD~209458.}. No match would produce a distribution of cross-correlation values consistent with random noise, i.e., with a flat line. The deviation from a flat line therefore indicates that a signal of some kind is detected. This is quantified by computing the $\chi^2$ of the cross correlation values around the planet radial velocity ($\chi^2_\mathrm{dir}$). The probability of measuring $\chi^2_\mathrm{dir}$ given $n$ degrees of freedom $P(\chi^2_\mathrm{dir},n)$ is computed and translated into a sigma value $\sigma_\mathrm{dir}$ for a Normal distribution with a two-tail test. A range of planet radial-velocity semi-amplitudes \kp\ within 4$\sigma$ from the literature value is explored;

\item The model is scaled to the observed planet/star flux ratio, Doppler shifted based on the same range of \kp\ indicated above, and removed from the data (i.e., injected with a scaling factor of $-1$).
Telluric lines are again removed and the residuals cross correlated with the model. A perfect match between the planet signal and the scaled model will result in a perfect subtraction. Therefore the cross-correlation function of the residuals will be consistent with random noise, which means its sigma value $\sigma_\mathrm{sub}$ computed as above from the chi square $\chi^2_\mathrm{sub}$ of cross-correlation values will be approximately zero;

\item The best-fitting model is found by maximizing the difference $\Delta\sigma=\sigma_\mathrm{dir}-\sigma_\mathrm{sub}$. In this study this is equivalent to maximizing the $\Delta\chi^2=\chi^2_\mathrm{dir}-\chi^2_\mathrm{sub}$, but this formulation is more general, as it allows us to test additional parameters when subtracting the model, e.g., line broadening due to rotation as in \citet{bro16}. It only requires adjusting the degrees of freedom $n$ when computing $P(\chi^2,n)$.
\end{itemize}

\subsection{Exploring the HDS parameter space}\label{lds_posterior}
The algorithm outlined in Section~\ref{sec:hds_metric} runs in approximately 2.5 minutes per model on a single core of a modern UNIX machine. It is therefore computationally too expensive to run a full MCMC for the HDS data that would include 10$^5$-10$^6$ models. However, since our final goal is to produce the {\sl joint} posterior distribution of LDS and HDS data combined, we can neglect regions of the parameter space which are already strongly disfavored by the LDS analysis. This is done by feeding the HDS analysis with a set of models sampled from the LDS posterior distribution of \citet{lin16}. The parametrization consists of molecular abundances for 7 species (CO, \water, \methane, CO$_2$, C$_2$H$_2$, NH$_3$, HCN) and five additional parameters to sample the temperature-pressure profile as in \citet{par14}. The 12-dimensional LDS posterior is sampled with 5,000 points, and the corresponding high-resolution, disk-integrated emission spectra are computed via line-by-line radiative-transfer calculations with the CHIMERA \citep{lin13, lin14, lin15} emission forward model. We assume a cloud-free atmosphere in this pilot study. Examples of LDS and HDS model spectra are shown in Figure~\ref{fig:lds_hds}.

We use the absorption cross-section database (and references there-in) from \citet{fre08} with subsequent upgrades described in \citet{fre14}.  The database comprises pre-computed cross-sections on a grid of temperature and pressure points sampled at intervals of 0.01~cm$^{-1}$. This corresponds to a resolution $R\sim\mbox{430,000}$, enough to resolve the individual lines at the CRIRES wavelengths over the physically relevant ranges of temperatures and pressures\footnote{We also tested higher-resolution (0.001~cm$^{-1}$) tabulated cross-sections computed with the HITRAN HAPI module and the HITEMP database, and found negligible differences with the 0.01-cm$^{-1}$ grid.}.  The opacity sources include H$_2$-H$_2$/He collision induced absorption, and molecular absorption due to the seven species listed above. Note these are the same opacities used for the LDS data retrieval in \citet{lin16}. 

\begin{figure}[ht!]
\plotone{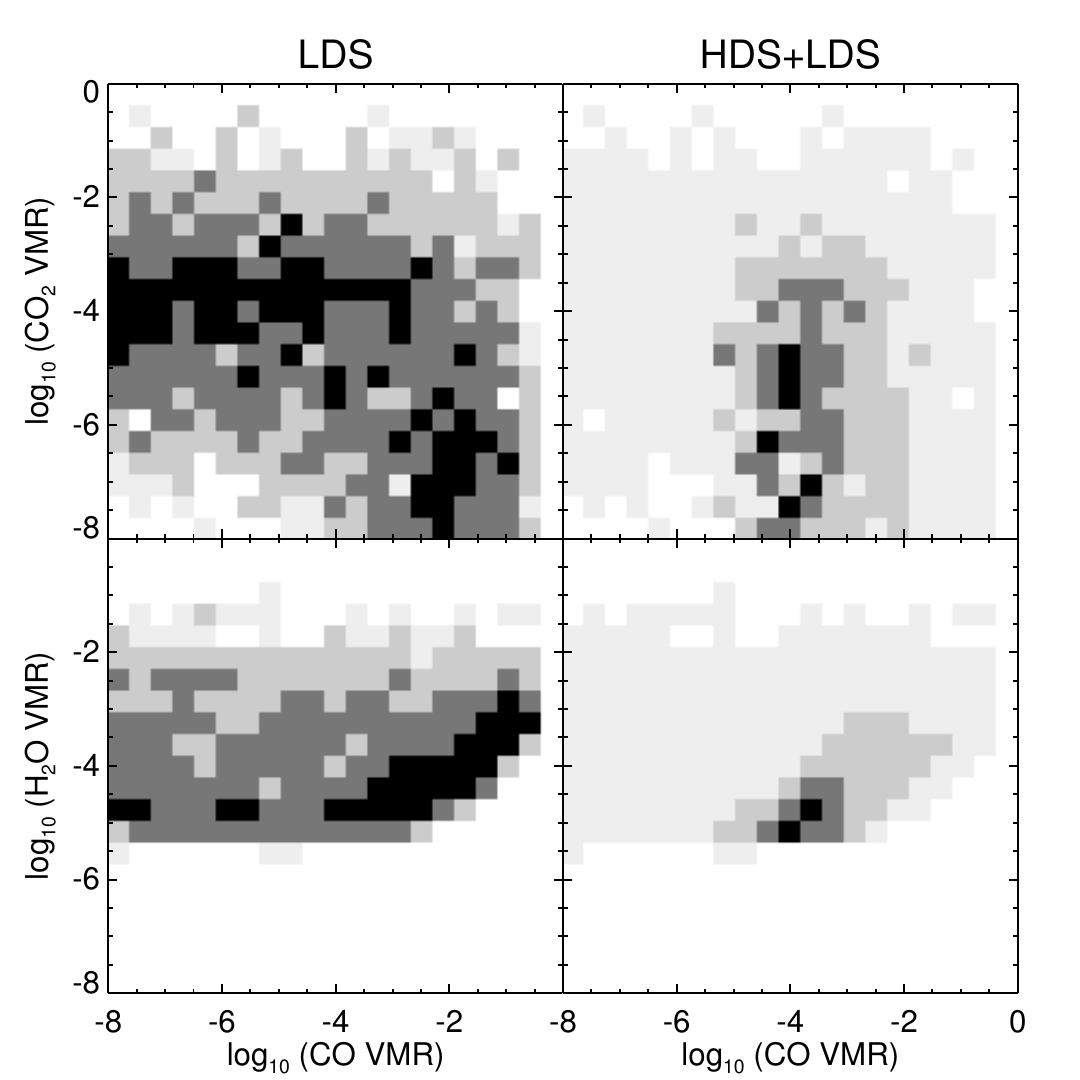}
\plotone{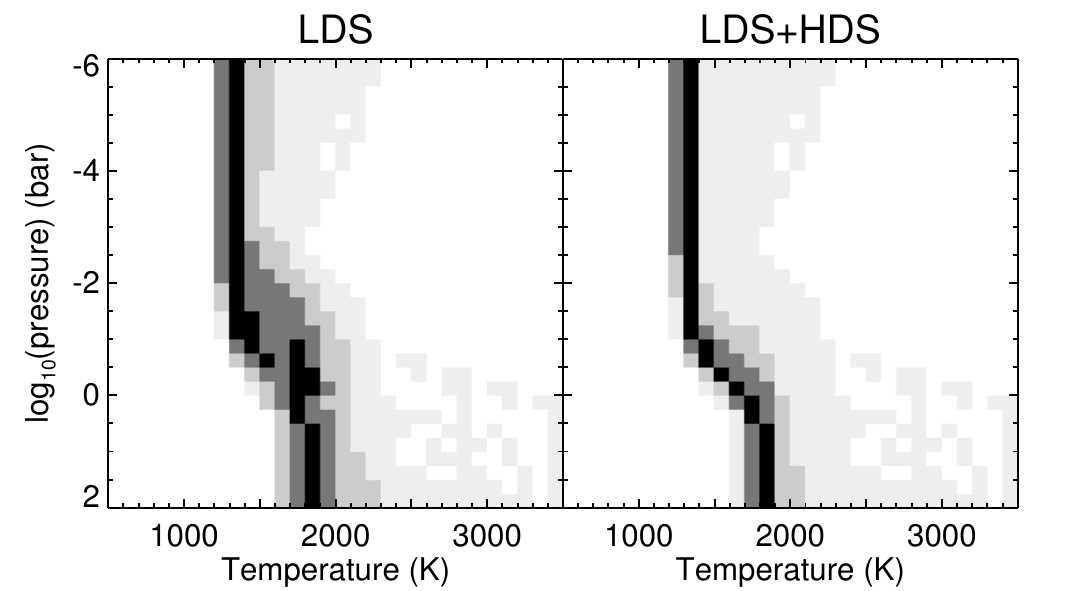}
\caption{Example of combined LDS and HDS analysis. {\bf Top:} The $\log$(abundance) of carbon monoxide is plotted against that of water vapor and carbon dioxide. Bottom: retrieved atmospheric pressure as function of temperature. The 1-$\sigma$, 2-$\sigma$, 3-$\sigma$, and $>$3$\sigma$ confidence intervals are plotted in greyscale (from black to light grey).}
\label{fig:co-h2o-co2_posterior}
\end{figure}

\subsection{Computing the joint posterior}
We compute the significance of the high-resolution models as explained in Section~\ref{sec:hds_metric}. To translate this information into an HDS posterior we bin the parameter space by 0.4 in $\log_{10}$(abundance). We then assume that in a posterior-sampling algorithm such as a MCMC, a model $n$-$\sigma$ deviant from the best-fitting model would be extracted with a probability $P=N(n)$, where $N$ denotes a Normal Distribution. As an example, the best-fitting model has $P=1$ and a model 2-$\sigma$ deviant $P=0.046$. The occupancy of a bin in the parameter space is therefore given by the sum of all the probabilities of the models falling in that range. The corresponding histogram is also computed for the LDS posterior. In this case the occupancy is given by the number of models falling in each bin as derived from the LDS retrieval analysis. 
In order to compute the joint posterior, we multiply the HDS and LDS posterior histograms, bin by bin. This is equivalent to multiplying the probabilities, which means we are treating LDS and HDS as two independent measurements of the same quantities. 
Figure~\ref{fig:co-h2o-co2_posterior} shows examples of the two-dimensional, LDS (left panels) and LDS+HDS (right panels) posteriors, obtained with the above method. The 1-, 2- and 3-$\sigma$ confidence intervals are obtained by normalizing the histogram by the total occupancy and cutting the cumulative density function at 61\%, 13.5\%, and 1.1\%, respectively. 

Since the HDS posterior is not obtained by freely exploring the parameter space, but it has been conditioned by the LDS analysis, our joint posterior is reliable only when the sampling is sufficiently dense to fill the whole 12-dimensional parameter space, i.e., within the 3-$\sigma$ confidence interval. This is why in Figure~\ref{fig:co-h2o-co2_posterior} we generically draw in light gray the region of the parameter space more than 3-$\sigma$ deviant from our best estimate of the parameter, without assigning any further confidence interval. In Section~\ref{sec:future} we further discuss the limits of this approach.

\section{Results}\label{sec:results}

The combination of low- and high-resolution spectroscopy of \pname\ improves constraints on both the vertical thermal structure and the molecular abundances of its atmosphere. 

\begin{figure*}[tb]
\epsscale{1.15}
\plotone{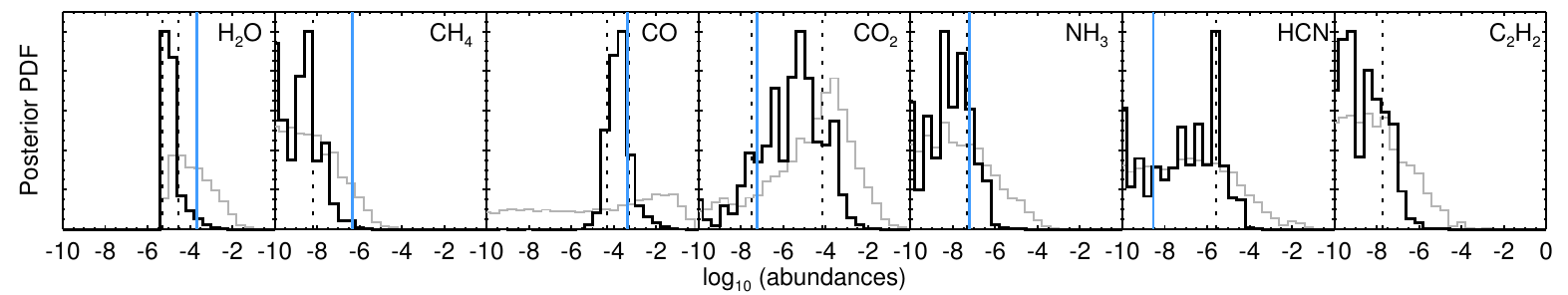}
\caption{Marginalized probability density function for the abundances of the seven species considered in this study. The LDS posterior (in grey) is compared to the joint LDS+HDS posterior (in black). The 15.9-84.1\% quantiles are shown with dotted lines, and the values for solar composition and chemical equilibrium with blue solid lines.}
\label{fig:abundances}
\end{figure*}

The bottom panel of Figure~\ref{fig:co-h2o-co2_posterior} shows that the confidence intervals for the retrieved temperature-pressure ($T$-$p$) profile shrink across the whole pressure range. The joint analysis points to a nearly-isothermal atmosphere, except between 0.1 and 10 bars where we measure a lapse rate of $\sim250$K per pressure decade.
%%Add isothermal?

\begin{figure}[t!]
\epsscale{1.1}
\plotone{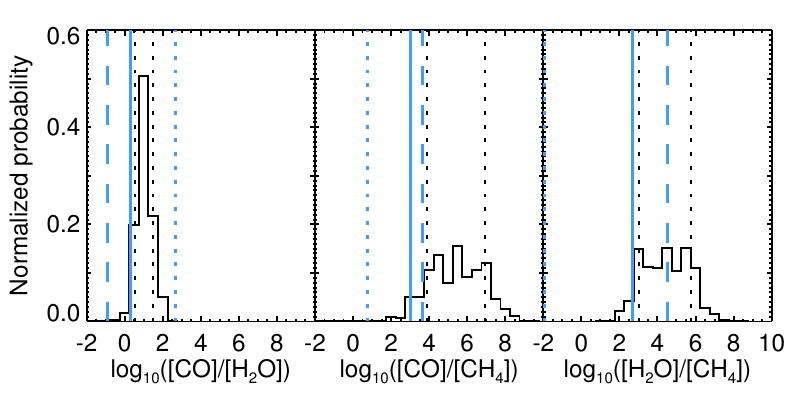}
\caption{Relative abundances of CO, CH$_4$, and \water\ derived from the joint LDS and HDS analysis. Black dotted lines denote the 1-$\sigma$ confidence intervals. Blue lines show the expected values for C/O=0.5 (solid), C/O=0.1 (dashed) and C/O=1 (dotted), (see Section~\ref{sec:results}).}
\label{fig:rel_abundances}
\end{figure}

Figure~\ref{fig:abundances} shows the marginalized joint posterior distribution of the 7 species included in our models (in black), compared to the posterior distribution from the low-resolution data only (in grey). 
The joint analysis significantly improves constraints on \water\ and CO ($\log_{10}$(VMR)$=[-4.97^{+0.42}_{-0.32},-3.80^{+0.51}_{-0.53}$], respectively). 
Their relative abundance is consequently tightly constrained to $\log$([CO]/[\water])~$=1.0^{+0.48}_{-0.44}$ (Figure~\ref{fig:rel_abundances}). Relative abundances with methane, which is non-detected, are instead looser as expected, with $\log_{10}$([CO]/[\methane])~$=5.4\pm1.5$ and $\log_{10}$([\water]/[\methane])~$=4.4^{+1.3}_{-1.4}$. 

\methane\ and NH$_3$ would be detectable in these CRIRES data if sufficiently abundant (VMRs $>10^{-5}$-$10^{-4}$). Their non-detection therefore tightens the upper limits on joint posterior abundances. 
The marginal improvement in the CO$_2$ posterior is unexpected. Except for abundances $>10^{-2}$, clearly excluded, CO$_2$ shows negligible spectral lines at 2.3~\micron. We indeed tested that removing CO$_2$ from the analysis does not alter the inference on the other species or the $T$-$p$ profile, which points to a non-detection in HDS data. However, CO and CO$_2$ are both opacity sources in the Spitzer 4.5 \micron\ channel. Their individual abundances cannot be easily disentangled with LDS data, as shown by the broad LDS posterior in Figure~\ref{fig:co-h2o-co2_posterior} (left-top panel). We suggest that the improved inference on CO when adding CRIRES spectra indirectly constrains CO$_2$ through the Spitzer IRAC measurement. Similar correlations within the LDS data are invoked to explain the small improvement in the joint posteriors of HCN and C$_2$H$_2$, negligible species at the wavelengths of these CRIRES observations.
If both CO and CO$_2$ are removed from the analysis, the joint posteriors only show a marginal improvement in \water\ abundance. This suggests that CO carries the large majority of the HDS signal, in line with the direct cross correlation described in Section~\ref{sec:obs}, but in tension with the clear detection of \water\ in the WFC3 passband.

We compare our measured abundances to the expectations for solar-composition and chemical equilibrium, at $p=0.1$ bar (representative of LDS spectra) and $T=1350$ K, including oxygen rainout due to enstatite/forsterite condensation. (Figure~\ref{fig:abundances}, blue solid lines). We find that \methane\ (at 2.6$\sigma$) and \water\ (at 1.9$\sigma$) are both under-abundant compared to the expectations.
 
We repeat the above calculations for a range of C/O and compute the expected relative abundances (Figure~\ref{fig:rel_abundances}, blue lines). [CO]/[\water] shows a clear (3.5$\sigma$) preference for C/O~$<1$, consistent with the stellar C/O of $0.46\pm0.05$ \citep{bre16a}. Relative abundances with \methane\ also constrain C/O~$<1$ at $>3\sigma$.

\begin{figure}[t]
\epsscale{1.2}
\plotone{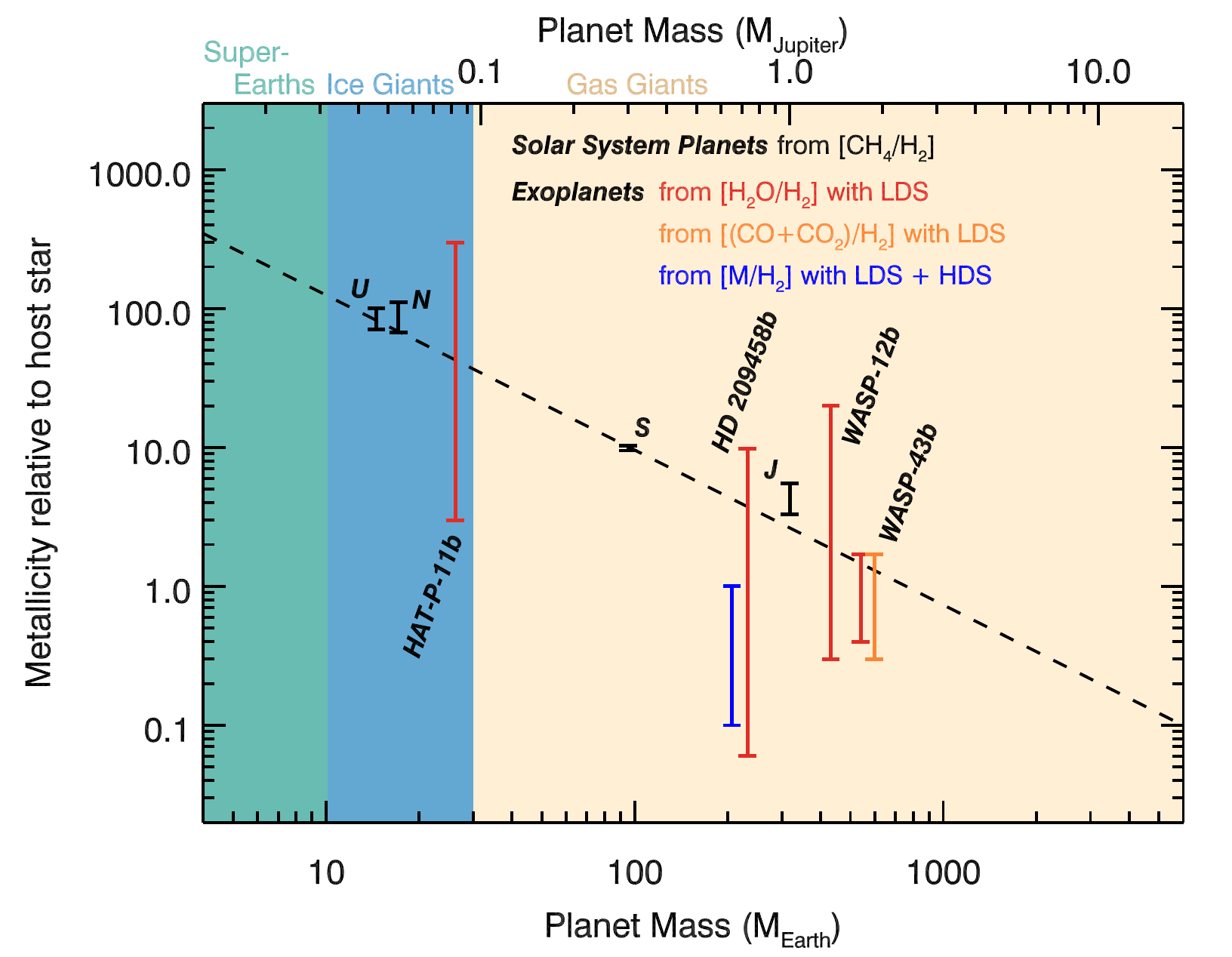}
\caption{Measured metallicity of solar system planets (in black) and exoplanets with available measurements (in colors), as a function of planet mass. The value for \pname\ obtained here is shown in blue. Exoplanet data from \citet{fra14, kre14, kre15, ste16}.}
\label{fig:mass-metallicity}
\end{figure}

We attempt a model-independent measurement of C/O and metallicity for the planet. We implicitly assume that molecular species not included in the analysis have negligible contribution to the planet spectrum. We compute the C/O by dividing the number densities of C and O atoms derived from the volume mixing ratios of molecular species. Although we obtain a tight estimate of C/O $=0.97^{+0.01}_{-0.03}$, we caution that oxygen rainout is likely to bias this measurement towards higher C/O.
We calculate a planet metallicity of $\log_{10}[$(M/H)/(M/H)$_\star]=-0.49^{+0.51}_{-0.48}$ (or 0.11-1.0$\times$ stellar, $1\sigma$) by assuming H$_2$/He = 0.193 and deriving the mole fractions of hydrogen and metals based on their stoichiometry. We adopt a stellar metallicity of (M/H)$_\star=9.7\times10^{-4}$ \citep{bre16b}.
\pname\ sits below the mass-metallicity relation for solar-system and extrasolar planets (Figure~\ref{fig:mass-metallicity}).

\section{Robustness against different molecular line databases and $T$-$p$ parametrization}
The completeness and accuracy of molecular line list databases can influence the template-matching technique utilized for HDS studies \citep[e.g.,][]{hoe15}. Here we explore the sensitivity of our joint retrieval to two different line lists for water vapor, namely the \citet{fre14} and the HITEMP2010 \citep{rot10} databases. 
On small wavelength ranges (10~nm), the two line lists produce models only mildly correlated (correlation of 0.35-0.4). Furthermore, their cross-correlation peaks a few km/s from zero lag, detectable at spectral resolutions above 2-3$\times10^4$. This indicates a mismatch in both line position and line strength potentially affecting the measured planet radial velocity, line broadening, and significance of high-resolution models. This could bias measurements of mass and orbital inclination of non-transiting planets, planet rotation, and the current analysis.
We therefore repeat the retrieval with spectra produced with the HITEMP2010 database. When \water\ is mixed with the other species and over the full wavelength coverage of the CRIRES observations (2.27-2.35~\micron) the agreement between the two analyses is remarkable, and we recover consistent best-fitting values for absolute and relative molecular abundances. We note, however, that since water is not confidently detected in these HDS data it is advisable to further investigate this aspect in future work.

We also ran this analysis with the $T$-$p$ profile parametrization of \citet{mad11}. This allows more flexibility in the upper atmosphere, where HDS data should be more sensitive. Although with larger error bars, we confirm the qualitative results of Section~\ref{sec:results} regarding abundances and C/O ratio. Compared to the previous parametrization, we retrieve a non-isothermal atmosphere across a wider range in pressure (10$^{-4}$-10 bar, with average lapse rate of $\sim160$K/decade). Furthermore, higher metallicity values (0.56-21$\times$ stellar) are allowed.

\section{Caveats and future improvements}\label{sec:future}
As the first attempt at combining LDS and HDS, this study necessarily simplifies some details. Firstly, the HDS analysis is conditioned on the posterior of the LDS analysis. Although this choice is consistent with our intent of computing a joint posterior distribution, it prevents us from exploring regions of the parameter space strongly disfavored by low-resolution spectra that could potentially be a good match to high-resolution spectra. 
Given the sample size of 5,000 points, we can reliably produce a joint posterior distribution for the 1-, 2- and 3-$\sigma$ confidence intervals, with the least significant regions of the parameter space possibly affected by the sampling. In the future we will explore algorithms capable of speeding up the comparison of model spectra to HDS data, so that a full MCMC will become feasible. In this way we will compute the LDS and HDS posteriors independently, which will fully highlight the strengths and weaknesses of both datasets, and the power of combining them.

Another aspect that is worth mentioning is that in the current implementation we are equally weighting the HDS and LDS data. This is supported by the similar evidence for CO+CO$_2$ (4.1$\sigma$) and \water\ (6.2$\sigma$) reported in \citet{lin16} and the detection of CO at S/N=5.2 in CRIRES data. Future work will be devoted to further assessing the relative weight of each dataset. This will enhance our predictive capabilities for designing coordinated observations with ground-based facilities and the James Webb Space Telescope. 

\acknowledgments
M.~B. acknowledges support by NASA through Hubble Fellowship grant HST-HF2-51336 awarded by the STScI, which is operated by the Association of Universities for Research in Astronomy, Inc., for NASA, under contract NAS5-26555. H.~S. acknowledges VICI grant 639.043.107, which is financed by The Netherlands Organisation for Scientific Research (NWO).
We thank R. de Kok for useful discussion on the inaccuracies of molecular line lists and line-by-line radiative-transfer modeling, C. Huitson and Z. Berta-Thompson for their input on the joint posterior distribution of LDS+HDS data, and J. Birkby, R. de Kok, and I. Snellen for extensively discussing the combination of low- and high-resolution spectroscopy.


%% for the paper.  Each keyword is check against the master list during
%% copy editing.  Individual instruments can be provided in parentheses,
%% after the keyword, but they are not verified.

%\vspace{5mm}
\facilities{HST(WFC3), Spitzer(IRAC), VLT(CRIRES)}

%\software{IRAF, cloudy, IDL}



%\appendix
%\section{Appendix information}
%\section{Author publication charges} \label{sec:pubcharge}


\listofchanges

\end{document}